\documentclass[leqno]{sigchi}
\usepackage{listings}
\usepackage{xparse}
\usepackage{outlines}
\usepackage{amsmath}
\usepackage{relsize}
\usepackage{etoolbox}
\usepackage{balance}       
\usepackage{graphics}      
\usepackage[T1]{fontenc}   
\usepackage{txfonts}
\usepackage{mathptmx}

\usepackage[math]{blindtext}
\usepackage[pdflang={en-US},pdftex]{hyperref}
\usepackage{color}
\usepackage{booktabs}
\usepackage{textcomp}
\usepackage{mathtools}
\usepackage{color,soul} 
\usepackage{changepage} 

\usepackage{microtype}        
\usepackage{ccicons}          
\usepackage{todonotes}
\usepackage{enumitem}

\makeatletter

\patchcmd{\env@cases}{\lbrace}{\@ldelim}{}{}
\makeatother

\newcommand{\code}[1] {\texttt{\textbf{#1}}}

\CopyrightYear{2020}
\setcopyright{acmlicensed}
\doi{https://doi.org/10.1145/3313831.3376593}
\isbn{978-1-4503-6708-0/20/04}
\conferenceinfo{CHI'20,}{April  25--30, 2020, Honolulu, HI, USA}
\acmPrice{\$15.00}

\def\plaintitle{Scout: Rapid Exploration of Interface Layout Alternatives through High-Level Design Constraints}

\def\emptyauthor{}
\def\plainkeywords{user interface design; wireframing; examples; alternatives; program synthesis; constraints}

\makeatletter
\def\url@leostyle{%
  \@ifundefined{selectfont}{
    \def\UrlFont{\sf}
  }{
    \def\UrlFont{\small\bf\ttfamily}
    
  }}
\makeatother
\newcommand{\quotateblock}[2]{%
\begin{adjustwidth}{0.5em}{0em}%
#1\footnotesize\textit{``#2''}%
\end{adjustwidth}%
}

\newcommand{\squeeze}[1]{\textls[-10]{#1}}
\newcommand{\squeezemore}[1]{\textls[-20]{#1}}

\def\pprw{8.5in}
\def\pprh{11in}

\definecolor{linkColor}{RGB}{6,125,233}
\hypersetup{%
  pdftitle={\plaintitle},
  pdfauthor={\emptyauthor},
  pdfkeywords={\plainkeywords},
  pdfdisplaydoctitle=true, 
  bookmarksnumbered,
  pdfstartview={FitH},
  colorlinks,
  citecolor=black,
  filecolor=black,
  linkcolor=black,
  urlcolor=linkColor,
  breaklinks=true,
  hypertexnames=false
}

\begin{document}

\title{\plaintitle}

\numberofauthors{2}
\author{%
Amanda Swearngin\textsuperscript{1}, Chenglong Wang\textsuperscript{1}, Alannah Oleson\textsuperscript{2}, James Fogarty\textsuperscript{1}, Amy J. Ko\textsuperscript{2}
  \alignauthor{
    \affaddr{Paul G. Allen School\textsuperscript{1}}\\
    \affaddr{University of Washington}\\
    \affaddr{Seattle, WA, 98195}\\
    \email{\{amaswea,clwang,jfogarty\}@cs.washington.edu}}\\
  \alignauthor{
    \affaddr{The Information School\textsuperscript{2}}\\
    \affaddr{University of Washington}\\
    \affaddr{Seattle, WA, 98195}\\
    \email{\{olesona,ajko\}@uw.edu}}\\
}

\maketitle

\begin{abstract}
\squeeze{Although exploring alternatives is fundamental to creating better
interface designs, current processes for creating alternatives are generally manual, limiting the alternatives a designer can explore.
We present Scout, a system that helps designers
rapidly explore alternatives through \mbox{mixed-initiative} interaction with high-level constraints and design feedback.}
Prior \mbox{constraint-based} layout systems use \mbox{low-level} spatial constraints and generally produce a single design. 
To~support designer exploration of alternatives, Scout introduces \mbox{high-level} constraints based on design concepts (e.g.,~semantic structure, emphasis, order) and formalizes them into \mbox{low-level} spatial constraints that a solver uses to generate potential layouts. In an evaluation with 18 interface designers, we found that Scout: (1) helps designers create more spatially diverse mobile interface layouts with similar quality to those created with a baseline tool and (2) can help designers avoid a linear design process and quickly ideate layouts they
do not believe they would have thought of on their own.
\end{abstract}

\ccsdesc[500]{Human-centered computing~Systems and tools for interaction design}

\keywords{Interface design, alternatives, program synthesis, constraints.}

\printccsdesc

\maketitle


\section{Introduction}
Alternatives are important in interface design. Studies have found that creating multiple designs in parallel results in higher-quality and and more diverse solutions \cite{buxton2007, dow2012}. 
\squeeze{When designers explicitly compare alternatives, it can enable them to make stronger critiques and better decisions \cite{dow2012a, tohidi2006}.} However, designers face many barriers in creating \mbox{high-quality} and diverse alternatives.
\squeeze{First, it is difficult to overcome fixation to think of completely new ideas \cite{Jansson1991}.}
Designers often sketch alternatives on paper \cite{buxton2007}, but such sketches can be difficult to change and a designer is still limited by the ability to envision new ideas to sketch. Example galleries \cite{herring2009examples} (e.g.,~Behance, Dribble) can help designers find inspiration from other design examples. However, a designer still needs to manually adapt examples into design alternatives. This may require \mbox{low-level} resizing, restyling, and relocating of interface elements. This can be particularly challenging for novice designers, as it requires knowledge of usability and visual design principles \mbox{\cite{LidwellWilliam2003Upod, Nielsen1990}} to maintain quality across alternatives. 

To aid designers in exploring and creating alternatives, we present Scout, a mixed-initiative system to help designers rapidly explore mobile interface layout alternatives. A designer can use Scout to express their interface elements and high-level constraints (e.g., semantic structure, order, emphasis), and Scout generates multiple alternative layouts satisfying those constraints to augment the designer's ideation. 

\squeeze{Scout applies constraint solving techniques to automatically generate alternatives.
Constraints have a rich history in interface design and visualization \mbox{\cite{borning2000constraintlayout,hottelier2014pbm,karsenty1993rockit,Xu2014beautification,zanden1991lapidary,Zeidler2013auckland}}.
However, such research has generally focused on reducing ambiguity in constraints to produce a single design.
In~contrast, our goal with Scout is to leverage a constraint solver to generate many diverse designs.
Additionally, \mbox{constraint-based} systems have generally focused on low-level spatial constraints (e.g.,~constraints expressed as mathematical equations in Apple Auto Layout~\cite{autoLayout}), which can be confusing and difficult for designers.
Scout lets designers specify high-level constraints based on usability and visual design principles like emphasis~\cite{emphasis2018} and clear hierarchies \cite{Kohler1967, white2011elements}, which Scout translates into \mbox{low-level} spatial constraints used by the underlying solver.} The key contributions of this work are: 
\vspace{-.25em}
\begin{itemize}
    \item Scout, a system to help designers rapidly visualize many layout alternatives for mobile interfaces through interaction with high-level constraints and feedback on alternatives.  
    \item A set of constraint encodings based on design principles, with solving algorithms that enable generating a range of diverse layouts for a set of interface elements.
    \item An evaluation with 18 interface designers, finding: (1)~that Scout can help them create more spatially diverse designs with similar quality to those created with paper and a baseline prototyping tool, and (2)~qualitative feedback demonstrating Scout's potential as a tool for early ideation and breaking out of a linear design process.   
\end{itemize}

\begin{figure*}[t!]
  \includegraphics[width=\textwidth]{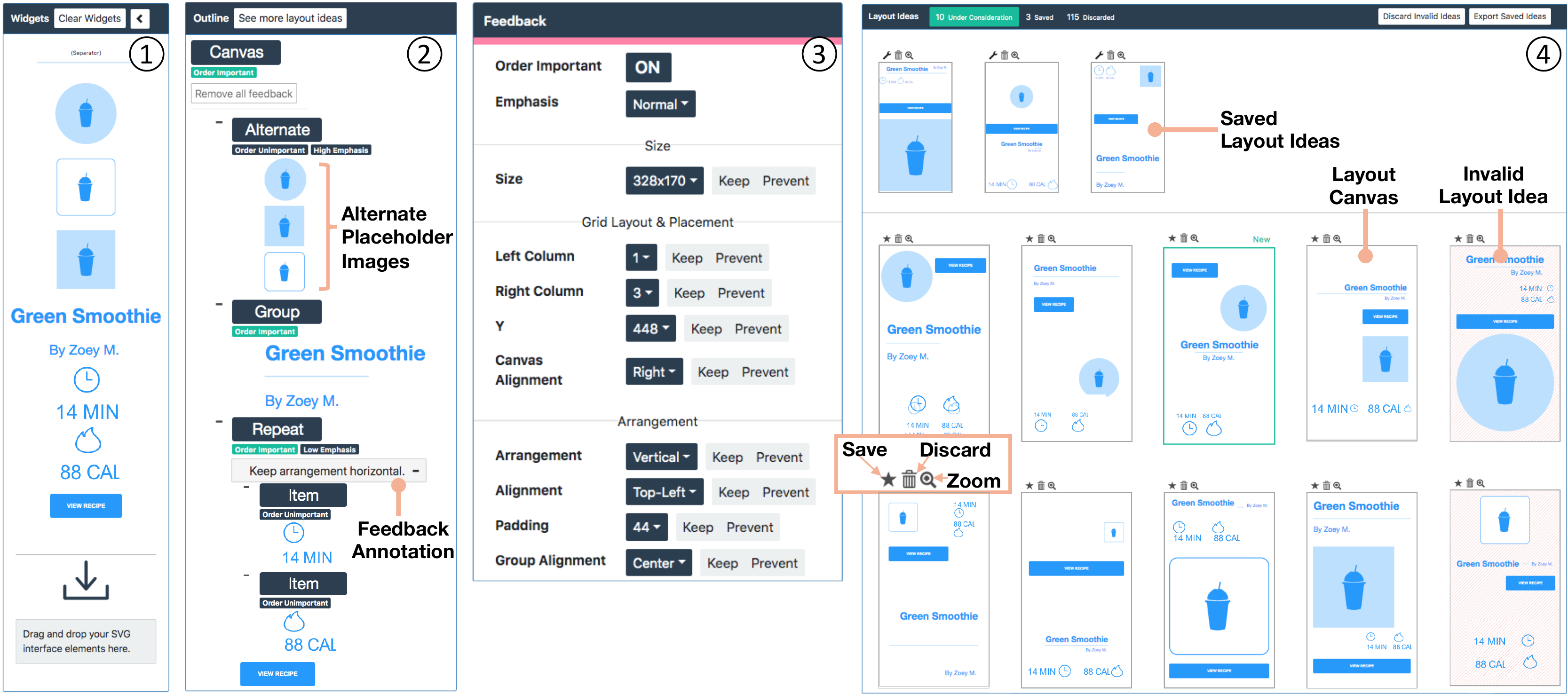}
  \caption{The Scout interface has four main panels: (1)~Designers import their interface elements by dragging their SVGs into the \emph{Widgets} panel. (2)~Designers create hierarchy and high-level constraints (e.g., grouping, order, emphasis) in the \emph{Outline} panel. (3)~Designers control generation of alternatives through the  \emph{Feedback} panel, which they can activate by clicking an element in the Outline panel or on an element in the Layout Ideas panel. (4)~The \emph{Layout Ideas} panel presents alternative layouts, which a designer can save, discard, or zoom in on.}~\label{fig:scout_interface}
  \vspace{-2em}
\end{figure*}
\section{Motivating Scenario}
To describe and motivate Scout, we consider an example scenario in which Eunice, a UX designer, is redesigning a recipe app landing page. Eunice has conducted a desirability study \cite{benedek2002measuring} of the current page. In such a study, people assign emotional and descriptive keywords to a design (e.g.,~``creative'', ``simple''). The top keywords assigned to Eunice's current design were ``dull'' and ``unrefined''. Eunice would like to change reactions to her landing page by using Scout to explore alternatives. First, Eunice imports a set of interface elements from her company's design library into Scout's Widgets panel (Figure~\ref{fig:scout_interface}.1). Then, Eunice clicks on elements in the widgets panel to add instances to her design's outline panel. Specifically, she adds 3 alternatives for a smoothie placeholder image, a header and subtext, calorie and time labels and icons, and a ``View Recipe'' button. Her~elements appear in Scout's outline panel (Figure \ref{fig:scout_interface}.2). 

\subsection{Specifying Hierarchy and High-Level Constraints}
Eunice next specifies high-level constraints on her elements. Scout lets designers group related elements, specify a relative order, and give elements high, normal, or low emphasis. We designed Scout's high-level constraints from common design principles for clear and usable layouts (e.g.,~\cite{constantine1999software,Kohler1967,LidwellWilliam2003Upod}). 

Eunice's first goal is to create a hierarchy. A key design principle is that interfaces should have a clear and organized hierarchy \cite{LidwellWilliam2003Upod}. Similarly, the structure principle \cite{constantine1999software} states that interfaces should keep related things together and unrelated things separate, motivated by Gestalt theory \cite{Kohler1967}. In Scout's Outline panel (Figure~\ref{fig:scout_interface}.2), Eunice creates a group for the ``Green Smoothie'' and ``By Zoey M.'' labels. For each group, Scout creates constraints to ensure these elements appear as visually distinct groups in layouts Scout generates. 

\squeeze{Eunice next wants to specify that the ``Green Smoothie'' label should always appear before the ``By Zoey M.'' label.} A~usability principle is that elements should appear in the order they are used for a task \cite{Nielsen1990}. \squeeze{Scout lets Eunice specify that order is \textit{important} or \textit{unimportant} for each group. When order is \textit{important}, Scout encodes a constraint to maintain the spatial reading order of grouped elements (i.e., left to right, top to bottom).} Scout also lets Eunice specify an element should appear first (e.g.,~a~header) or last (e.g.,~a~footer) in a layout. 

\squeeze{Many interfaces include repeating patterns of elements (e.g.,~a~list, a~grid). Scout supports \textit{repeat groups} to ensure the layout of subgroups is consistent. Eunice creates a repeat group for the calories and minutes labels and icons (Figure~\ref{fig:scout_interface}.2).
When Scout generates layouts (Figure~\ref{fig:scout_interface}.3), it keeps the layout of the subgroups of elements that a designer places in the group consistent (i.e., alignment, arrangement, order, padding). Scout also infers repeating patterns of elements within a group to suggest when this constraint can be applied. }

Finally, Eunice wants to see layouts that use alternate versions of the smoothie image placeholder, so she creates an \textit{alternate group} with 3 different placeholder images (Figure~\ref{fig:scout_interface}.2, ``Alternate''). When Scout creates layouts, it uses only one of the three placeholders in each layout. 

Eunice has created her high-level constraints, so she clicks ``See more layout ideas'' at the top of Scout's Outline panel (Figure~\ref{fig:scout_interface}.2). Scout displays a set of 20 layouts satisfying Eunice's high-level constraints in the Layout Ideas panel (Figure~\ref{fig:scout_interface}.4). Eunice sees that some layouts show the smoothie image too small and the calorie icon pairs too large in relation to other elements. She decides to set \textit{emphasis} levels for these elements. Emphasis is an interface design principle \cite{white2011elements}, stating that interfaces should have a main focal point to let a person know what to do next \cite{emphasis2018}. Scout allows specifying \textit{High}, \textit{Normal}, or \textit{Low} emphasis.  Eunice uses the Feedback Panel to give the smoothie placeholder \textit{High Emphasis} and the minutes and calories repeat group \textit{Low Emphasis}. Scout will then adjust the size and position of her elements to make them more or less visually prominent. 

\begin{figure}[t!]
\centering
  \includegraphics[width=\columnwidth]{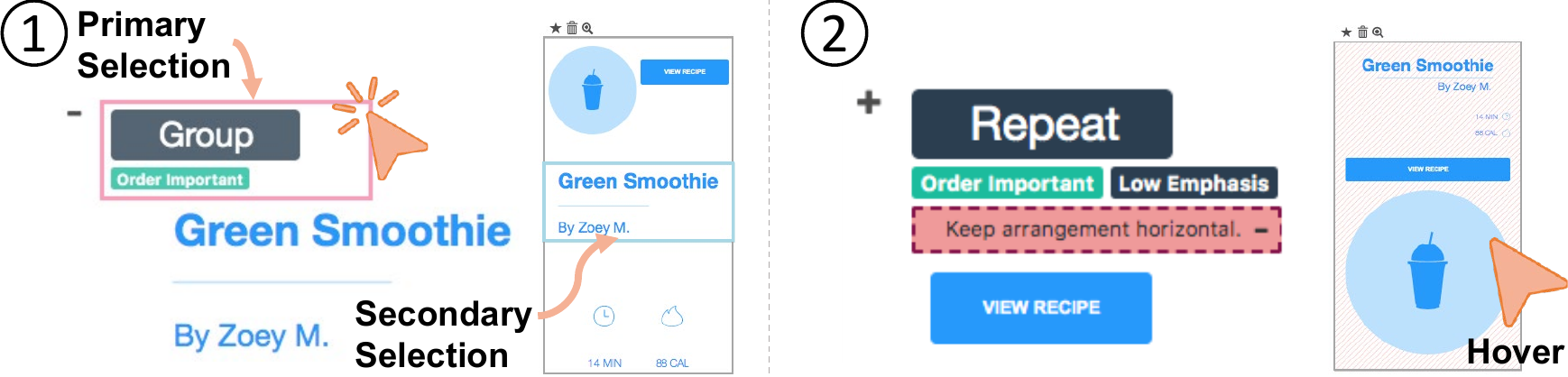}
  \caption{(1) Designers can click nodes in Scout's Outline panel to make them the primary selection, which highlights corresponding elements in each layout in the Layout Ideas panel. (2) Designers can hover over a layout, and Scout highlights conflicting feedback annotations. }~\label{fig:scout_conflict}
  \vspace{-2em}
\end{figure}
\subsection{Feedback \& Layout Curation}
After reviewing Scout's generated layouts, Eunice decides to use a horizontal layout for the minutes and calories group. She clicks a layout with a horizontal group, and Scout displays a pink outline around the selected element (Figure~\ref{fig:scout_conflict}.1) to indicate the Feedback panel is active for that element. The Feedback panel displays feedback properties that let Eunice ``Keep'' or ``Prevent'' specific property values in the alternatives (e.g., ``Keep alignment left''). Eunice clicks the ``Keep'' button next to the arrangement dropdown to tell Scout to use a horizontal arrangement for the group in future alternatives. Eunice's feedback appears in Scout's Outline panel as a \textit{feedback annotation} (i.e.,~``Keep arrangement horizontal''). Eunice can also activate the Feedback panel by clicking an element in the Outline panel. In that case, Scout will set each feedback property dropdown to ``Vary'' until a ``Keep'' or ``Prevent'' feedback is applied. Scout supports multiple ``Keep'' and ``Prevent'' values for a property (e.g.,~``Keep arrangement horizontal OR vertical''). Scout lets Eunice give several types of feedback, including on the top-level canvas (e.g.,~``Keep layout grid 4 columns''), groups (e.g.,~``Keep arrangement horizontal''), and elements (e.g.,~``Keep location here'').

Values that a designer ``Keeps'' or ``Prevents'' can cause a conflict in existing layouts. Eunice sees that Scout has put red diagonal stripes over two layouts. She hovers her mouse over one of the layouts, and Scout highlights the conflicting feedback, "Keep arrangement horizontal", in red (Figure~\ref{fig:scout_conflict}.2). This layout has a conflict because the minutes and calories repeat group is vertical, and not horizontal. 
\squeeze{When Scout detects a conflict, Scout tries to repair the layout to match the designer's feedback. If it cannot repair the layout, Scout retrieves a new layout to replace it, ensuring that Eunice's Layout Ideas panel is continually filled with new layouts as she applies her feedback.}

\begin{figure}[b!]
\centering
  \includegraphics[width=\columnwidth]{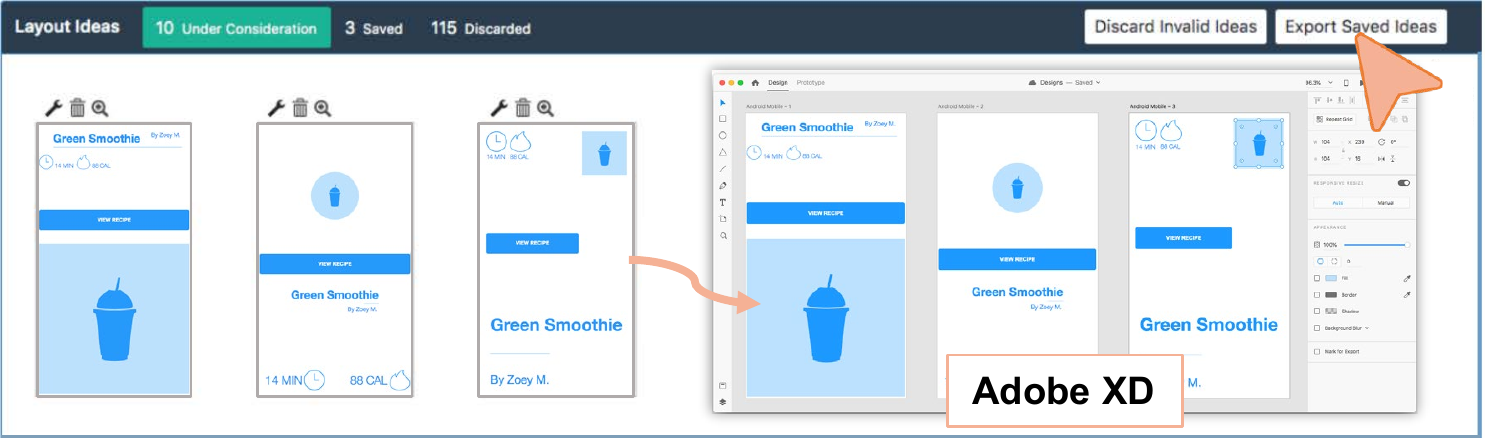}
  \caption{Designers can export their saved layout ideas to an SVg to import into a prototyping tool like Adobe XD.}~\label{fig:export_to_xd}
  \vspace{-2em}
\end{figure}
Using Scout, Eunice explores over 100 layouts.  She discards several by clicking the \textit{trashcan} icon above each layout. As she finds layouts she likes, she saves them for export by clicking the \textit{star} icon above each layout. Scout pins these to the top of the Layout Ideas panel (Figure~\ref{fig:scout_interface}.4). After Eunice has found 3 diverse layouts, she decides to refine them by exporting them out of Scout to edit in her favorite interface design tool (Figure~\ref{fig:export_to_xd}). Scout exports each layout as an SVG with editable shapes and properties. Using Adobe~XD, Eunice adjusts the alignment and relative size of the layouts until she feels they are ready for further feedback from her colleagues. 
\lstdefinestyle{code}{%
   basicstyle=\ttfamily
}

\begin{figure*}
\centering
  \includegraphics[width=\textwidth]{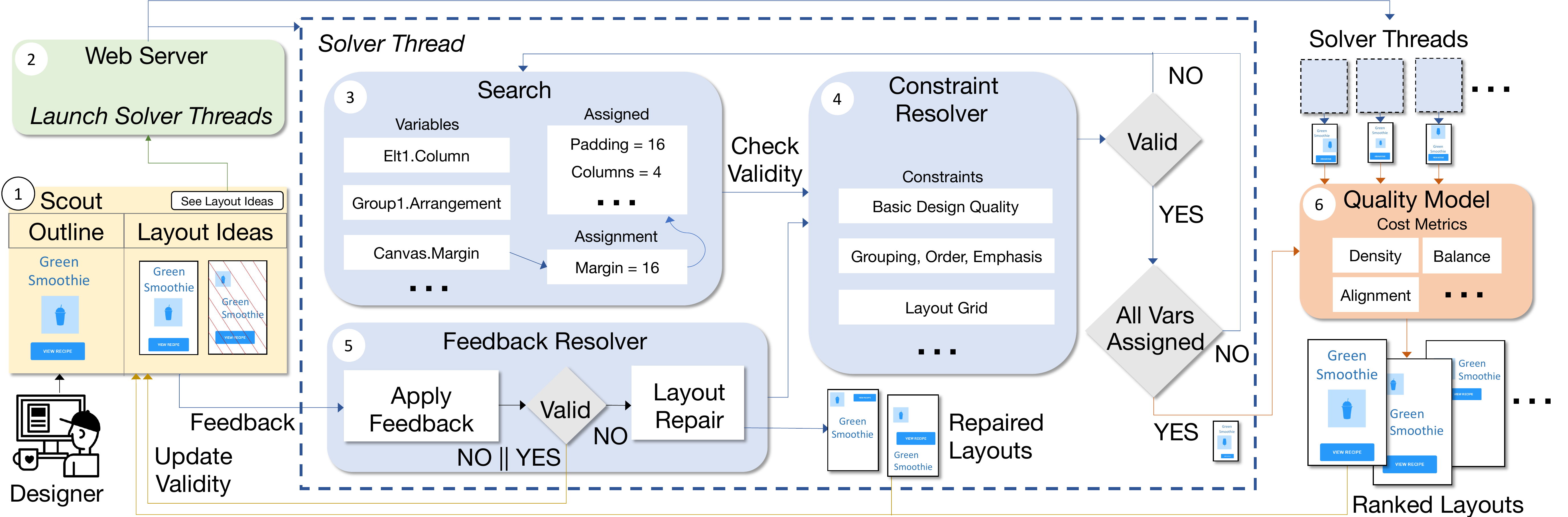}
  \caption{\squeeze{Scout System Overview: (1)~A~designer gives input to Scout via an outline of interface elements and feedback on layout alternatives. (2)~A~web server generates layouts by launching multiple solver threads. (3)~Each solver thread searches over variable assignments. (4)~A~constraint resolver checks the assignments against constraints. (5)~A~feedback resolver applies designer feedback and repairs layouts. (6)~A~quality model ranks resulting layouts.}
  }~\label{fig:architecture}
  \vspace{-2em}
\end{figure*}

\section{Scout Architecture and Implementation}
\squeezemore{Before developing the current version of Scout, we conducted informal interviews with 6 UX designers, including their use and feedback on an early version of Scout. We draw upon their insights to support several key system design choices, including:~(1)~prioritizing interactive performance, (2)~improving design quality through a design quality ranking model and utilizing a layout grid, and (3)~providing stability in a designer's current set of designs through a feedback resolver which can repair designs that conflict with new designer feedback.}

Figure~\ref{fig:architecture} illustrates Scout's architecture. A designer provides a set of interface elements, each as an SVG (Figure~\ref{fig:architecture}.1). When the designer requests new layouts, Scout sends their interface elements and high-level constraints to the server (Figure~\ref{fig:architecture}.2), which launches multiple solver threads to generate layouts with interactive performance  (Figure~\ref{fig:architecture}.3-5). Each thread produces a layout, consisting of an \textit{x position}, \textit{y position}, \textit{width}, and \textit{height} for each element. Scout ranks each layout by a score computed with a quality model (Figure~\ref{fig:architecture}.6) based on design quality metrics (e.g., alignment, balance). Scout then displays the ranked set of layouts to the designer visually as an SVG \textit{layout canvas}. A designer can give feedback on layouts, and a feedback resolver (Figure ~\ref{fig:architecture}.5) applies the feedback and attempts to repair conflicting layouts. 

\subsection{Generating a Layout Alternative}
Scout generates layouts through a modified branch and bound search \cite{russell2002artificial}, which generates a satisfying set of variable assignments (e.g., \textit{alignment}, \textit{arrangement}) (Figure~\ref{fig:architecture}.3) with respect to a set of design and high-level constraints on interface elements (Figure~\ref{fig:architecture}.4). Each variable has a domain of values Scout can assign through its search (e.g., \textit{alignment} is one of top, left, x-center, y-center, bottom, right). Each constraint is a formalized as an equation encoded into the Z3~\cite{de2008z3} constraint solver operating on one or more variables (e.g., element \textit{size} and \textit{position}). Throughout this paper, we format \code{constraint names} in a typewriter font and \textit{variable names} in italics. The next section details Scout's constraints and variables. 

\squeeze{Figure~\ref{fig:architecture} shows Scout's process to generate a layout. First, Scout's search process (Figure~\ref{fig:architecture}.3) generates a single variable assignment for an element or group. The constraint resolver (Figure~\ref{fig:architecture}.4) then uses the Z3~\cite{de2008z3} constraint solver to determine whether the assignment is valid}. The constraint resolver translates \mbox{high-level} constraints specified by designers into formalized \mbox{low-level} variables and constraints on interface elements and layout behavior, which we detail in a later section. If the assignment is not valid, Scout backtracks in the search and reassigns the variable. If the assignment is valid and other  variables remain unassigned, Scout assigns another variable and checks it through the constraint resolver. Finally, when Scout has assigned all variables, it produces a layout with a position and size for each interface element.

\squeeze{To create spatially diverse layouts, Scout randomizes assignment order of variables and values using a uniform distribution. After Scout produces a layout, it encodes a constraint that prevents that same layout from appearing again. If a solver thread cannot produce a layout, Scout discards that thread. Scout can be configured to launch a variable number of threads based on system capabilities. For our evaluation, we configured Scout to launch 20 solver threads each time the designer requests new layouts. On the machine we used (Ubuntu~18.10 with AMD Ryzen~7~1800x processor, 8~cores x 16~threads, 32~GB memory), Scout typically returned 15 layouts per request, containing 9 elements each, in less than 5 seconds. Such resources are common for many designers (e.g.,~who also work with image and video data), but Scout could also run in a configuration with solver threads shifted to a scalable cloud service. Scout currently supports a complexity typical of many mobile interfaces, but future research would likely be needed to scale Scout's algorithms to larger and more complex interfaces.} 

\subsection{Ranking Layouts by Quality Metrics}
\squeeze{Scout's layout search space is extremely large (i.e.,~trillions). Some layouts are not well-aligned or visually pleasing, and designers need a way to prioritize higher-quality layouts during their exploration. We created a quality model (Figure~\ref{fig:architecture}.6) to compute a \textit{layout quality} score for each layout, formalizing some design principles described in interface design literature.}
Scout uses these scores to rank higher-scored layouts toward the top of the Layout Ideas panel. We adapted this model from  \cite{riegler2018measuring}, which presents a set of metrics to computationally measure mobile interface complexity (e.g.,~misalignment, imbalance, density). For each group of elements $G=\{e_1,\dots,e_n\}$ in a layout, Scout computes $\mathit{quality_g}(G)$ which considers element size, balance, and alignment.
$$\mathit{quality_g}(G)=s_\mathit{size}(G) + s_\mathit{balance}(G) + s_\mathit{alignment}(G)$$
The \textit{size} score penalizes groups with excessively large or small elements. It computes the sum of the normalized width and height of each element (i.e., normalized by the width and height of the canvas), divided by the number of elements. 
$$s_\mathit{size}(G)=\frac{1}{2|G|}\sum_{e\in G}(\frac{e.\mathit{w}}{W} + \frac{e.\mathit{h}}{H})$$
The \textit{balance} score rewards groups with \mbox{evenly-spaced} margins between consecutive pairs of elements. It computes the difference between the average horizontal and vertical margins $G.\mathit{avg\_margin_h}$,
$G.\mathit{avg\_margin_v}$ and the maximum horizontal and vertical margins $G.\mathit{max\_margin_h}$, $G.\mathit{max\_margin_v}$. 
$$s_\mathit{balance}(G)= \frac{1}{2}(\frac{G.\mathit{avg\_margin_h}}{G.\mathit{max\_margin_h}} + \frac{G.\mathit{avg\_margin_v}}{G.\mathit{max\_margin_v}})$$
The \textit{alignment} score measures quality of alignment within a group. For each pair of elements $e_1,e_2$, $\mathit{NumAlignment}$ returns the number of horizontal (i.e.,~top, \mbox{y-center}, bottom) and vertical (i.e.,~left, \mbox{x-center}, right) alignments between those elements.
\squeeze{$\mathit{NumPossibleAlignments}$ returns the maximum number of alignments the two elements could have. For~example, if $e_1$ and $e_2$ are horizontally arranged and have the same height, they can have a maximum of 3 alignments (i.e.,~top, \mbox{y-center}, bottom). The score therefore measures the proportion of alignment pairs out of the total number of alignments.}
$$s_\mathit{align}(G)=\frac{1}{|G|}\sum_{(e_i,e_j)\in G,\ i \neq j}\frac{\mathit{NumAlignment}(e_i,e_j)}{\mathit{NumPossibleAlignment}(e_i,e_j)}$$
Finally, Scout computes an overall \emph{layout quality} score as a weighted-average of each group quality score $\mathit{quality_g}(G)$, where each group is weighted by its area. The \emph{layout quality} score also includes: (1) a density score $s_\mathit{d}$ to measure the ratio of the entire layout area covered by elements, and (2) a \emph{group quality} score treating the top-level set of groups as an additional group (i.e., to measure the quality of layout of those top-level groups on the canvas). 
$$\mathit{quality_l}(L)=\frac{\sum\limits_{G\in L}G.\mathit{area}\cdot\mathit{quality_g}(G)}{\sum\limits_{G\in L} G.\mathit{area}} + s_d + \mathit{quality_g}(L)$$    




\subsection{Feedback \& Layout Repair}
After Scout produces an initial set of layouts, a designer can update the outline (i.e., change the grouping, emphasis, or order of elements) or give feedback on variables (i.e.,~canvas, group, or element variables), prompting Scout's feedback resolver (Figure~\ref{fig:architecture}.5) to recheck the validity of each layout. For any ``Keep'' feedback, Scout encodes an equality constraint into the solver (e.g.,~$group\_arrangement == ``vertical''$). Conversely, ``Prevent'' feedback is encoded as an inequality constraint (e.g.,~$group\_arrangement \neq ``vertical''$). Scout checks validity of each layout with respect to the outline and constraints, then updates their validity in the interface (i.e.,~with red diagonal stripes over invalid layouts). Scout uses Z3's \cite{de2008z3} \code{unsat core} to obtain the smallest set of constraint clauses that cannot be satisfied. When a designer hovers over an invalid layout, Scout examines these conflicting clauses and highlights the corresponding feedback annotations.

A designer can apply feedback that makes many layouts invalid. To prevent the designer from needing to frequently request new layouts, Scout continuously generates layouts as a designer applies feedback. To minimize disruption to the current set of layouts, Scout tries to return similar layouts through a layout repair module (Figure~\ref{fig:architecture}.5). Layout repair iteratively selects variable assignments to remove from an invalid layout from z3's~\cite{de2008z3} \code{unsat core} until it becomes valid (Figure~\ref{fig:architecture}.4, Layout Repair). To prevent overwhelming a designer with too many layouts, Scout does not repair or generate new layouts if the Layout Ideas panel has more than 50 valid layouts (i.e., the number that could reasonably be visible on a 24-inch monitor). However, a designer can still request new layouts with the ``See more layout ideas'' button. 

\subsection{Constraint Encodings \& Design Variables}\label{design_variables}
Scout generates layouts through an assignment of concrete values to a set of variables, allowing it to explore many combinations of element arrangement, alignment, position, and size. Scout defines \textit{canvas} variables (e.g., layout grid, margin, baseline grid), \textit{group} variables (e.g., arrangement, alignment, padding), and \textit{element} variables for \textit{position} (e.g.,~x,~y), and \textit{size} (e.g., width, height). Each variable has a domain, curated from design guidelines \cite{materialDesign} and layout design literature \cite{babich2017grids}, together with constraints defining its behavior. Scout uses these constraints, a designer's high-level constraints, and basic design quality constraints, to check the validity of a layout (Figure~\ref{fig:architecture}.4). We include formalized equations for all constraints in our supplementary materials. 

\lstset{
 breaklines=true
}

\subsubsection{Ensuring Basic Design Quality}
Scout encodes three \code{basic design quality} constraints for every layout, an approach also used by Beilik~et.~al. \cite{bielik2018robust} in encoding a set of ``Robustness Properties'' for Android layouts. For each element, Scout enforces a \code{stay-in-bounds} constraint that keeps elements inside the bounds of the layout canvas.  Scout also encodes a pairwise \mbox{\code{non-overlapping}} constraint on the bounding boxes of each pair of elements. Finally, Scout encodes \code{minimum sizing} constraints for each element from design guidelines (e.g.,~touch targets should be at least 48x48 pixels \cite{materialDesign}). 

\subsubsection{Placing Elements on the Layout Canvas}
\squeeze{Scout uses a layout grid to place elements on a canvas by encoding constraints on an element's bounding box. A~\textit{layout grid} is a common method designers use to place elements, which can improve alignment, consistency, and visual organization \cite{vinh2010ordering}. It consists of \textit{margins} (i.e.,~spacing on the outside of the canvas that all elements must be placed inside), \textit{columns} (i.e.,~vertical containers for placing elements on the canvas), and \textit{gutters} (i.e.,~spacing between columns where elements must not be placed). Mobile interfaces typically use a 2 to 4 column layout grid \cite{materialDesign}, within which elements or groups must begin and end on a column and not in a gutter, and can span multiple columns. Scout defines 4 layout grid variables for a canvas: \textit{margin}, \textit{columns}, \textit{gutter width}, and \textit{column width}. 
Based on these variables' values, Scout encodes \code{layout grid} constraints requiring the left and right edges of elements and groups that are direct children of the canvas to begin and end on the edge of a column.}

Baseline grids define the vertical spacing of a design, aid horizontal alignment, and create hierarchy \cite{babich2017grids}. They consist of horizontal lines at even intervals to which all components should align. Scout defines a \textit{baseline grid} variable that allows designers to examine different baseline grid options. Based on this, Scout encodes \code{baseline grid} constraints specifying that elements have 
a \textit{y position} aligned to a baseline grid line
and a \textit{height} that is a multiple of the \code{baseline grid} value.  

\subsubsection{Resizing Elements}
To explore different element sizes, Scout defines a \textit{size} variable for each element with a domain of the form $(width, height, sizing\_factor)$. $sizing\_factor$ is used to enforce consistent resizing within groups. Scout \mbox{pre-computes} \textit{width} and \textit{height} domains using two strategies: \textit{maintain aspect ratio} and \textit{increase width}.
For both strategies, Scout computes a set of $(width, height, sizing\_factor)$ triples along baseline grid increments (i.e., \code{4px}), where \textit{width} values range from a minimum determined by element type to the canvas width (e.g., \squeeze{\code{[(20,20,1), (24,24,2), ...]}}). For~\textit{maintain aspect ratio} elements (e.g., images, icons), \textit{height} values vary from a minimum for each element to the canvas size. For~\textit{increase width}, height values do not vary (e.g., \squeeze{\code{[(120,40,1), (124,40,2), ...]}}). Scout encodes each pre-computed set of triples as the domain to a \textit{size} variable. This is a performance optimization because Z3 does not efficiently compute multiplication constraints (i.e.,~otherwise needed for maintaining an aspect ratio).

\subsubsection{Grouping and Order}
Designers can group elements in the Outline panel to keep them together. Scout varies layout of grouped elements based on three variables: \mbox{\textit{alignment},} \textit{arrangement}, and \textit{padding}. Scout encodes constraints aligning grouped elements along 6 possible \textit{alignment axes}: left, x-center, right, top, y-center, and bottom. Scout currently aligns all elements within a group to a single axis. Scout defines four \textit{arrangement} domain values for each group: horizontal, vertical, balanced rows, and balanced columns. Each \code{arrangement} constraint encodes rules based on the position and size of grouped elements. Scout defines \code{padding} constraints that work with \code{arrangement} constraints to add spacing between grouped elements while keeping them relatively close to each other. Finally, Scout defines \code{visual hierarchy} constraints to keep the within-group \textit{padding} smaller than the group's distance to other groups in the layout to visually separate them.

To allow designers to control element order, Scout lets designers specify order of groups as \textit{important} or \textit{unimportant}. For groups with \textit{important} order, Scout encodes an \code{ordering} constraint that combines with \code{arrangement} constraints to keep the elements in the fixed order specified in the outline. For groups with \textit{unimportant} order, Scout encodes a constraint on the height and width of the group bounding box, according to the \textit{arrangement} variable. This allows elements to change position if other constraints are met (e.g., horizontal arrangement). If order is \textit{important} for the top-level canvas, Scout encodes a constraint on each pair of elements such that the bottom edge of an element must be equal to or above any element later in the ordering. Scout also lets designers specify that an element should be \textit{first} or \textit{last} in a group, which enables specifying a fixed position for elements like a label, header, or footer. Scout encodes constraints requiring these elements to be first or last in the group. For the top-level canvas, Scout encodes pairwise constraints stating the top edge of a \textit{first} element should be above all other elements and the bottom edge of a  \textit{last} element should be below all other elements. 

\subsubsection{Emphasis}
To support designers specifying a visual hierarchy, Scout includes emphasis constraints based on design guidelines \cite{white2011elements} that state emphasis can be increased or decreased by modifying an element's size in relation to other elements and position in the reading order. Scout supports 3 levels of emphasis levels: \textit{low}, \textit{normal}, and \textit{high}. All~elements have \textit{normal} emphasis by default. For elements with \textit{low} or \textit{high} emphasis, Scout encodes a \code{size decrease only} or \code{size increase only} constraint on the element's \code{size} variable that allows the element's area to decrease or increase from its original area. Scout~also specifies a \code{relative size} constraint stating that (1) elements with \textit{high} emphasis should have a larger \textit{height} or \textit{width} than elements without high emphasis, and (2) elements with \textit{high} emphasis should either have a larger \textit{area} or appear earlier in the order than elements without \textit{high} emphasis.  
Scout encodes similar \code{low emphasis} constraints, constraining \textit{low} emphasis elements to have a smaller \textit{height} or \textit{width} than elements without \textit{low} emphasis, and to either have a smaller \textit{area} or appear after elements without \textit{low} emphasis. 

\subsubsection{Alternate Representations and Repeating Patterns}
Alternate groups let a designer show alternate elements (i.e.,~SVGs) in different layouts. For each alternate group, Scout creates a \textit{representation} variable with a domain corresponding to the elements the designer groups. Scout's search (Figure~\ref{fig:architecture}.3) assigns a value to this variable, which a designer could ``Keep'' or ``Prevent'' like other variables.

Repeat groups indicate a layout should be kept consistent across multiple subgroups (e.g., a list, a grid). A repeat group contains a set of subgroups, each with the same number of elements, the same types (e.g., button, text, image), in the same order. Each element in a subgroup has a corresponding element in all other subgroups, determined by their order. Figure~\ref{fig:scout_interface}.2 shows a repeat group containing two pairs of icon and label (i.e., minutes and calories labels with corresponding icons that should always be arranged similarly). For each subgroup in a repeat group, Scout encodes a constraint that requires the \textit{arrangement}, \textit{alignment}, \textit{padding}, and \textit{order} variable values of all subgroups to be equal. Scout also encodes a constraint requiring any increase or decrease in the \textit{size} variable be the same for corresponding elements in each subgroup.
\section{Evaluation}
\begin{figure}
\centering
  \includegraphics[width=\columnwidth]{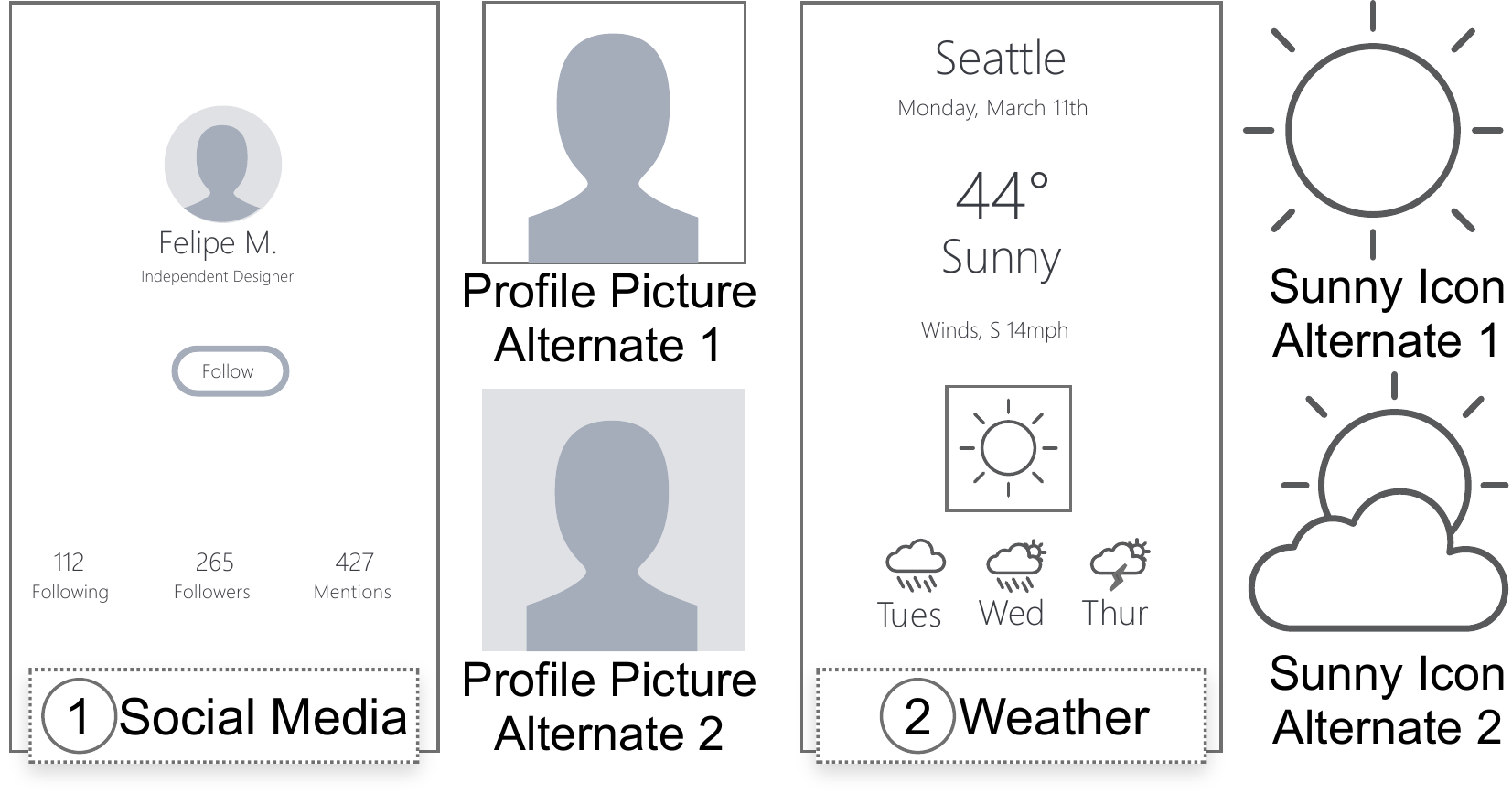}
  \caption{\squeezemore{Our provided components for the \textit{Social Media} and \textit{Weather} scenarios, including alternate images for the profile picture and sunny icon.}}~\label{fig:task_components}
  \vspace{-2em}
\end{figure}

To understand Scout's benefits and limitations, and to examine how different designers might use \mbox{mixed-initiative} layout ideation, we conducted a \mbox{within-subjects}, \mbox{mixed-methods} evaluation centered on three research questions:
\vspace{-.25em}
\begin{itemize}
    \setlength{\itemsep}{0pt}
   \item RQ1: \squeezemore{Does Scout help designers of varying expertise generate} \textbf{more diverse} interface layouts than with a baseline tool? 
   \item RQ2: \squeezemore{Does Scout help designers of varying expertise generate} \textbf{higher quality} interface layouts than with a baseline tool?
   \item RQ3: How does Scout \textbf{affect designer processes} of exploring potential interface layouts?
\end{itemize}

\subsection{Participants}
We recruited 18 interface designers (5M, 13F, ages \mbox{18-32}), 9 in each of 2 \textit{Experience Level} groups:
(1)~\textit{Professional Designers}, with >=1 year of professional UI/UX design experience; and (2)~\textit{Non-Professional Designers}, who had built at least one complete interface prototype but had <1 year of professional experience. \textit{Professional} designers reported a range of professional experience (1~to~3 years of experience:~5; \mbox{3 to 5} years:~2; more than 5 years:~2). 
Five \textit{Non-Professional} designers self-reported no professional UI/UX experience, while four reported less than 1 year of experience.

\subsection{Procedure}
Each participant completed two wireframe prototyping tasks, varying \textit{Interface} to use \textit{Scout} and a \textit{Baseline} prototyping tool, Adobe XD. To better examine the use of Scout, rather than participant learning of Scout's interface, participants completed a 20-minute Scout tutorial and warmup task (i.e.,~exploring layouts for a To Do List). After the tutorial and before proceeding with the task, participants demonstrated how to use Scout's grouping, alternate, and repeat group constraints. All participants had experience with Adobe~XD and had used similar tools (e.g.,~Figma, Sketch), so~we did not include a corresponding \textit{Baseline} warmup task. We~collected screen and audio recordings and notes while participants completed tasks, then interviewed them after each task to reflect on their process using each \textit{Interface} and on differences in using Scout versus their current process. 

We developed two \textit{Scenario} as a hypothetical setting for participant tasks: redesigning two app screens for a design agency: (1)~a \textit{Social Media} profile screen (Figure~\ref{fig:task_components}.1), and (2) a \textit{Weather} app screen (Figure~\ref{fig:task_components}.2). We selected familiar screen types so designers could focus on improving screen layout rather than the content. As in the scenario with Eunice described earlier, our task scenarios described that the agency had conducted a desirability study \cite{benedek2002measuring} and that keywords assigned to the weather and social media app screens were ``dull'' and ``familiar''. We asked designers to redesign for the keywords ``clean'' (i.e.,~"uncluttered and well-aligned") and ``compelling'' (i.e.,~"has a clear point of emphasis"), attributes of good layouts from design guidelines~\cite{emphasis2018, LidwellWilliam2003Upod, white2011elements}. 

\begin{figure}
\centering
  \includegraphics[width=\columnwidth]{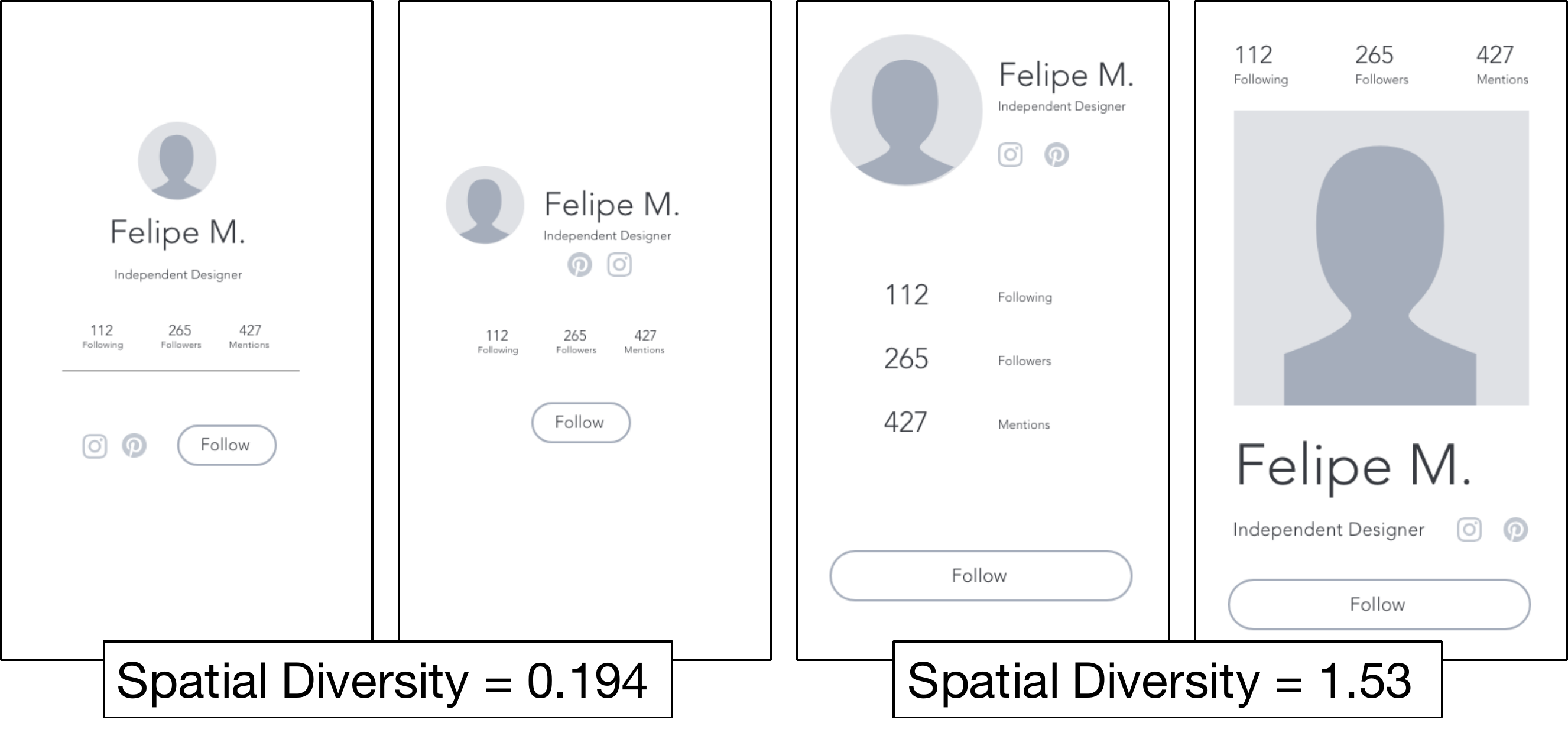}
  \caption{\squeezemore{To illustrate our spatial diversity score, the least diverse~(left) and most diverse~(right) pairs of participant-produced \textit{Social Media} layouts.}}~\label{fig:diversity_scores}
  \vspace{-2em}
\end{figure}

We had the designers create three diverse layouts as alternative redesigns of the \textit{Social Media} and \textit{Weather} screens. We provided pre-created wireframe elements for the original design, 2 alternate profile pictures for \textit{Social Media}, and 2 alternate sunny icons for \textit{Weather} (Figure~\ref{fig:task_components}). The task content encouraged using Scout's \textit{repeat group} (e.g.,~for the days of the week and weather icons), but we did not require the designers to use any particular constraint. The app screens were similar in complexity (e.g.,~number of elements, groups of elements). 
Because Scout is focused on layout, we limited designer use of Adobe XD to spatial features (e.g.,~position, size, font size) and not non-spatial features (e.g.,~color, font type). We required the designers to use only the provided elements without overlapping or rotating them.  

Each designer completed a 30 minute task for each of 2 conditions: \textit{Baseline} and \textit{Scout}.  We allowed sketching on paper in both conditions. In \textit{Baseline}, the designers could use the time to sketch and create alternatives. For \textit{Scout}, the designers used Scout for 20 minutes, saved 3 layouts, and spent 10 minutes refining and finalizing the layouts in Adobe XD. We interviewed the designers after each task, and at the end of the study. The total session time was less than 2 hours per designer. To address learning or other carryover effects, we counterbalanced \textit{Interface} (i.e.,~\textit{Scout} or \textit{Baseline}) and \textit{Scenario} (i.e.,~\textit{Social Media} or \textit{Weather}) using a Latin square design. We performed our analysis using mixed effect models, treating \textit{Participant} as a random effect and modeling \textit{Interface}, \textit{Scenario}, and \textit{Experience Level} as fixed effects.  

\subsection{Results} 
\begin{table}
\begin{center}
\begin{tabular}{{c|c|c|c|c|c|c}}
          & G & RG & AG & O & E & FB \\
\toprule
Prop. (n=180)& 20\% & 9\% & 9\% & 24\% & 21\% & 17\% \\
\% Des. (n=18) & 94\% & 78\% & 89\% & 94\% & 83\% & 72\% \\
\end{tabular}
\end{center}
\caption{The proportions of high-level constraints of each type specified by designers after the Scout task, and the percentage of designers who specified each type of high-level constraint (i.e., group (G), repeat group (RG), alternate group (AG), order (O), emphasis (E), feedback (FB).} ~\label{tab:hlc_summary}
  \vspace{-2em}
\end{table}
Overall, the designers generated an average of 97 layouts during the Scout task (min: 19, max: 280, SD: 81). At the end of the Scout task, designers had an average of 10 high-level constraints specified (min: 6, max: 17: SD: 3.8). Table~\ref{tab:hlc_summary} summarizes the percentage of designers that used each type of high-level constraint and the proportions of each type of high-level constraint specified at the end of the study.

\textit{RQ1: Does Scout help designers of varying expertise generate \textbf{more diverse} interface layouts than with a baseline tool?}\\
We wanted to understand Scout's impact on helping designers explore more diverse layouts. Given Scout's focus on spatial arrangement, we defined diversity as \textit{spatial diversity}. Although there are existing computer vision dissimilarity metrics \cite{miniukovich2015visual}, they are not suitable to compare the wireframes from our study (i.e.,~the fact that wireframes are primarily whitespace causes these approaches to fail). We instead developed a spatial diversity score to estimate the effort needed to adapt one layout to another (i.e,~transformation distance \cite{hahn2003similarity}). Gajos~et.~al.~\cite{gajos2005cross} present a dissimiliarity metric capturing transformation distance by comparing each layout along a set of dimensions (e.g.,~orientation, representation). We adopted a similar approach, defining \textit{spatial diversity} for a pair of layouts containing the same elements in terms of 3 metrics: (1)~mean \textit{position change} computes the mean of the distance that each element moved between the two layouts, (2)~mean \textit{size change} computes the mean of how much the area of each element changed between the two layouts, and (3)~mean \textit{relational distance change} ($s\_rel$) measures the mean of the position change of an element in relation to all other elements in the layout, computed as follows, where $\mathit{dist}(e_i,e_j)$ calculates the distance between the centers of  two elements.  
$$s\_rel(L, L')=\frac{2}{n(n-1)}\cdot\sum_{1\le i< j\le n}|\mathit{dist}(e_i,e_j)-\mathit{dist}(e_i',e_j')|$$
Overall spatial diversity was a weighted sum of mean \textit{position change}, mean \textit{size change}, and mean \textit{relational distance change}. To weight the metrics, we divide by the maximum value of that metric for any pair of elements across all layouts in our evaluation (i.e., normalizing metrics into the range $[0,1]$). Figure~\ref{fig:diversity_scores} shows two pairs of designs from our study with the smallest and largest spatial diversity scores.

\begin{figure}[t!]
\centering
  \includegraphics[width=\columnwidth]{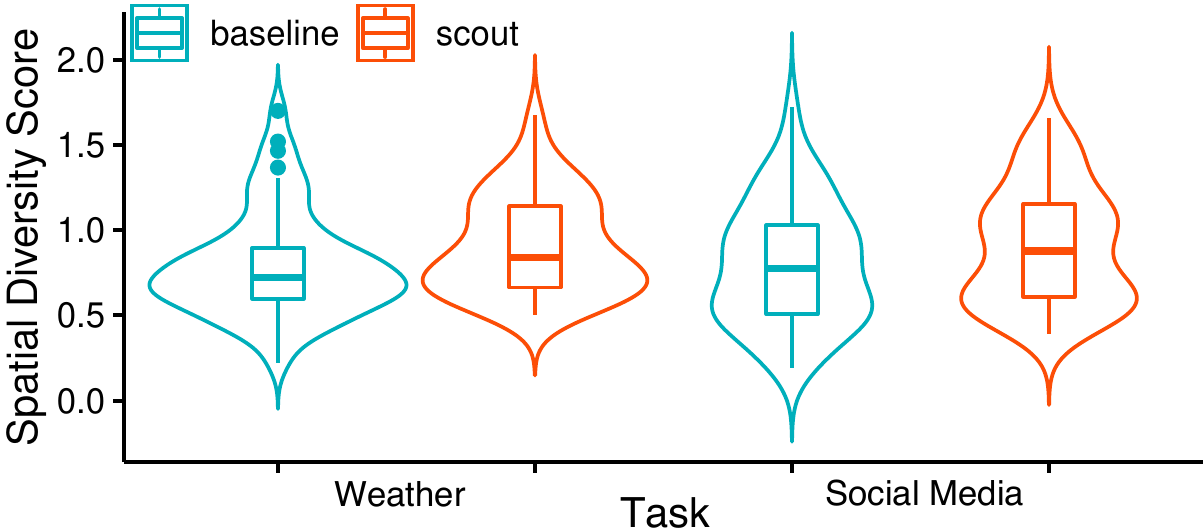}
  \caption{Violin plots of the spatial diversity scores for each set of pairs by a designer within an Interface/Scenario combination. Scout layouts had higher spatial diversity for both scenarios.} ~\label{fig:within_designer_diversity}
  \vspace{-2em}
\end{figure}

\squeeze{To examine Scout's impact on spatial diversity within designs created by individual designers, we conducted a \mbox{within-designer} analysis. We computed spatial diversity for each pair of designs created by a designer with each \textit{Interface}, excluding the original design (i.e., 3 pairs per designer per \textit{Interface}).} \squeeze{Figure~\ref{fig:within_designer_diversity} shows violin plots by \textit{Interface} and \textit{Scenario}. The 54 pairs of \textit{Scout} designs were 12 percent more spatially diverse ($M = 0.880, SD=0.290$) than the 54 \textit{Baseline} pairs ($M = 0.788, SD = 0.356$).} Spatial diversity scores were not normally distributed, so we conducted an \mbox{aligned-rank-transform} analysis \cite{wobbrock2011aligned}, which indicated a significant effect of \textit{Interface} on spatial diversity ($F_{1,86} = 5.05$, $p < 0.027, d=0.435$). The analysis did not find a significant effect of \textit{Experience Level} on spatial diversity score ($F_{1,14} = 0.009$, $p < 0.926$). 

To examine whether Scout helped designers create layouts that were more spatially diverse relative to the original design, we computed a spatial diversity score for each layout relative to the original (i.e., 3 pairs per designer per \textit{Interface}). Scout helped designers create layouts that were 15 percent more spatially different ($F_{1,86} = 5.35, p < 0.023, d=0.45$) than the original design ($Scout - M = 0.926, SD = 0.343$, $Baseline: M = 0.807, SD = 0.315$). Although the effect of \textit{Experience Level} on spatial diversity was not significant ($F_{1,14} = 0.038, p < 0.848$), our analysis showed a significant interaction effect between \textit{Interface} and \textit{Experience Level} ($F_{1,86} = 4.46 , p < 0.038$). Using Scout increased spatial diversity by 35\% for \textit{Non-Professional} participants ($Baseline: 0.749, Scout: 1.01$), while decreasing spatial diversity for \textit{Professional} participants by 2 percent ($Baseline: 0.866, Scout: 0.847$). An interaction contrast, corrected with Holm's sequential Bonferroni procedure, indicated this difference when using Scout according to \textit{Experience Level} was significant ($\chi^{2}(1, n = 27) = 4.46 , p < 0.035, d=0.41$). 

\begin{figure}[t!]
\centering
  \includegraphics[width=\columnwidth]{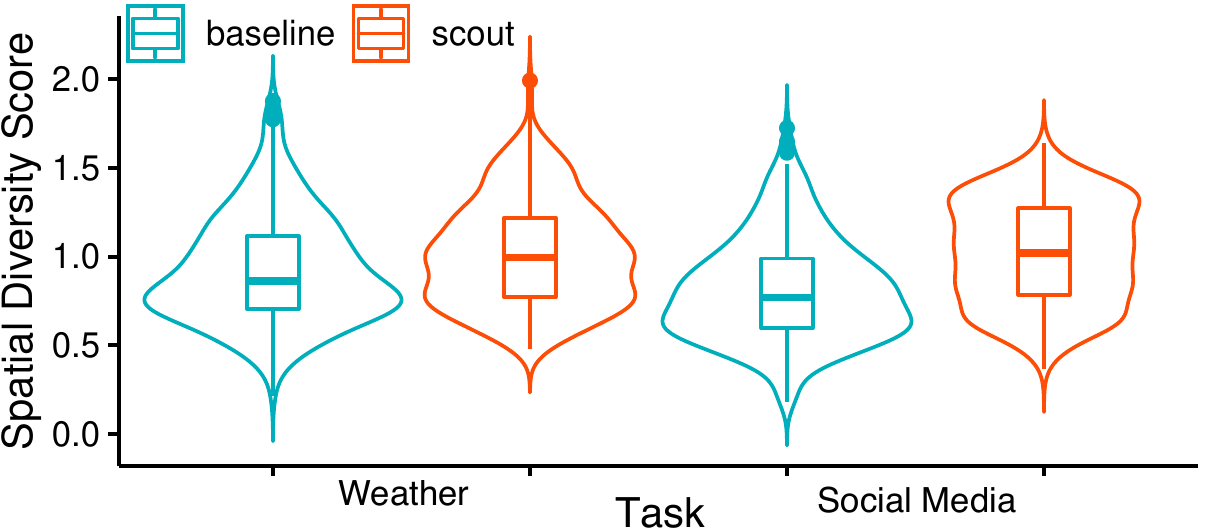}
  \caption{Violin plots of spatial diversity scores across all design pairs by all designers within a Interface/Scenario combination. Scout layouts had higher overall spatial diversity for both scenarios.} ~\label{fig:across_designer_diversity}
  \vspace{-2em}
\end{figure}

\begin{table}[b!]
\vspace{-1em}
\begin{tabular}{{c|c|c|c|c|c|c}}
         \textbf{n = 54}  & VB & TH & E & A & W & LQ \\
\toprule
Scout (M) &  2.67 & 3.07 & 2.39 & 2.5 & 2.82 & 5.37 \\
Scout (Std) & 0.97 & 0.84 & 1.09 & 0.82 & 0.87 & 1.0 \\
Baseline (M) & 3.09 & 3.01 & 2.65 & 2.83 & 2.77 & 5.73 \\
Baseline (Std) & 0.96 & 0.79 & 1.13 & 0.88 & 0.98 & 1.24 \\
\end{tabular}
\caption{Summary statistics of the sum of the two quality scores awarded by expert evaluators to designers' layouts from our user study, including visual balance (VB), typographical hierarchy (TH), emphasis (E), alignment (A), whitespace (W), and overall layout quality (LQ).} ~\label{tab:quality_scores}
\vspace{-2em}
\end{table}
Finally, we examined Scout's effect on overall spatial diversity across designers. We computed the entire set of pairwise spatial diversity scores within the \textit{Social Media} and \textit{Weather} scenarios. Figure~\ref{fig:across_designer_diversity} shows Scout increased the overall mean spatial diversity score for the \textit{Social Media} scenario by 26 percent ($Baseline: M = 0.811, SD = 0.290, n = 351$, $Scout: M = 1.02, SD = 0.289, n = 361$) and for the \textit{Weather} scenario by 10 percent ($Baseline: M = 0.926, SD = 0.296, n = 351$, $Scout: M = 1.02, SD = 0.306, n = 351$). Spatial diversity scores were not normally distributed (\mbox{Shapiro-Wilk} $W > 0.974, p < .0001$). Using the Wilcoxon rank sum test, we found a significant difference in means for both the \textit{Social Media} ($W = 50640, p < .0001, r=0.342$) and \textit{Weather} ($W = 37243, p < .0001, r=0.154$) scenarios. 

\textit{RQ2: Does Scout help designers of varying expertise generate \textbf{higher quality} interface layouts than with a baseline tool?}
We assessed the quality of participant designs with a panel of 2 independent interface designers, each with at least 3 years of professional UX design experience, who each evaluated each design on a layout evaluation rubric.  
The rubric included 5 items. The first 3 focused a design being ``compelling'': (1)~\code{visual balance} - \textit{``The layout is easy to scan, and all elements are aligned with respect to axes of symmetry''}, (2)~\code{typographical hierarchy} - \textit{``All elements follow a typographical hierarchy and are easily readable and proportionally sized with respect to each other.''}, and (3)~\code{clear point of emphasis} - \textit{``The wireframe has a clear point of entry or a single visually salient feature, that does not overwhelm the design.''}. The final 2 focused on a design being ``clean'': (4)~\code{alignment} - \textit{``All elements in the wireframe are aligned with one or more other elements.''}, and (5)~\code{whitespace} - \textit{``Whitespace effectively used to separate unrelated components.''} The designers score each rubric item as ``Great'' (2 points), ``Good'' (1 point), or ``Needs Improvement'' (0 points).  For each design, we computed an overall \textit{layout\_quality} score which weights the ``compelling'' scores (i.e., visual balance (vb), typographical hierarchy (th), and emphasis (e)) and ``clean'' scores (i.e., alignment (a) and whitespace (w)) equally, summed across the two designers (d). 
$$\mathit{layout\_quality}(L) =\sum_{d\in D}\frac{vb + th + e}{3} + \frac{a + w}{2}$$
Overall, the \textit{Scout} designs had slightly lower \textit{layout\_quality} scores (M = 5.37, SD = 1.0, n = 54) than the \textit{Baseline} (M = 5.73, SD = 1.24, n = 54). The layout quality scores were not normally distributed, so we assessed their significance using an aligned-rank-transform analysis \cite{wobbrock2011aligned} which indicated the difference in the mean \textit{layout\_quality} score was not significant ($F_{1,87} = 2.35$, $p < 0.13$). Table~\ref{tab:quality_scores} shows the results for individual quality rubric items, suggesting that Scout designs can be improved for visual balance, emphasis, and alignment. 
\textit{RQ3: How does Scout \textbf{affect designer processes} of exploring potential interface layouts?}\\
We asked designers to reflect on their process to explore alternative layouts. After each task, we conducted semi-structured interviews asking designers to describe their strategy to develop diverse, compelling, and clean designs. We conducted an additional \mbox{semi-structured} interview at the end of the session, where designers compared their experiences using each tool and discussed how they might or might not use a tool like Scout in their design process. To analyze this data, two of the researchers collaboratively conducted a qualitative inductive content analysis \cite{patton_qualitative_2014} on the interviewer's notes, with a sensitizing concept of \textit{differences across design processes} when using Scout versus using other tools. 


\textbf{Designers viewed Scout layouts as compelling over clean.}\\
8 designers felt their Scout designs were more ``compelling'' than their Baseline designs, while 4 designers thought their Baseline designs were more compelling. 
However, ``compelling'' to the designers did not necessarily mean having a clear point of entry and clean hierarchy (i.e., as defined in the task).
Some designers interpreted ``compelling'' as ``interesting'', or ``atypical'', like P4:
\vspace{-4pt}
\quotateblock{}{It does a good job with the compelling thing...The hierarchy is not dull or boring or and to some extent is not even familiar. ... Like this [Scout design], it breaks [design] cliches, that's for sure. It does a good job of not being boring...}
\vspace{-1pt}

Conversely, when comparing their Baseline and Scout designs, 10 designers felt they produced ``cleaner'' designs in the Baseline (5 designers thought their Scout designs were cleaner). P20 noted that making clean designs with Scout might take more work:
\vspace{-4pt}
\quotateblock{}{You'd have to put a lot of rules on it to get it as clean as you'd want it to be. For example, this [Scout design I made] is not very clean-looking, but I could picture moving it around a bit, and it would be clean...}
\vspace{-1pt}

\squeeze{\textbf{Designers followed a ``mix and match'' process with Scout.}}\\
When discussing the strategy they followed to explore layouts in Scout, 11 designers mentioned a ``mix and match'' practice where they observed some portion of a Scout layout that they liked, then either (1) combined the portion with part of one or more other layouts during the ``Refinement'' phase, or (2) used feedback to ``Keep'' one or more properties of the layout to see how it would pair with alternatives for the remaining layout. Designers also mentioned an iterative strategy of looking through the initial set of layouts and giving feedback (5 designers) or adjusting the high-level constraints (5 designers) based on what they saw in the set of layouts. 9~designers also mentioned that Scout was useful to visualize many combinations of element layouts. In seeing many alternate arrangements, P7 found it useful to have both effective and ineffective layouts to choose from.
\vspace{-4pt}
\quotateblock{}{It stretches your understanding of what's possible. They were wide and broad and messy, and they draw attention to why they don't work... You can look at it more as this is close but we need to change something a little bit to make it better.}
\vspace{-1pt}



\textbf{Designers followed a less linear process with Scout.}\\
Scout may have also helped designers follow a less linear process of creating alternatives. When discussing their Baseline task process, 12 designers mentioned a linear process of looking at a design and thinking about how to change it into a new design. In contrast, only 2 designers mentioned a linear design process in Scout. A few designers mentioned Scout could help them resist focusing in on a few designs too early, and as a consequence, explore more divergent ideas. 

\textbf{Scout can help designers think of new ideas.}\\ Nine designers mentioned that Scout helped them think of new ideas they might not have had on their own.  P2 mentioned struggling to create three diverse designs in the Baseline task:
\vspace{-4pt}
\quotateblock{}{I feel like I was able to get two really good designs, but the middle one I really don't like. I wasn't able to come up with a third one...I feel like I probably just needed more time...}
\vspace{-1pt}

After the Scout task, P2 described Scout helping them create a design that they would not otherwise have come up with:
\vspace{-4pt}
\quotateblock{}{[T]he way Scout helped with that was, I wasn't even looking to make something like the third layout. I wouldn't have thought to put the image at the bottom of the page, this gigantic one.}
\vspace{-1pt}

11 designers mentioned that some or all of their Scout layouts were different than a typical weather or social media profile screen. 
This was desirable for the study because we asked them to create diverse layouts. 
desirable. However, 2 designers noted that breaking design conventions might not always be desirable, and said that they tend to prioritize familiarity more to prevent from distracting the user.

\textbf{Designers would use Scout during layout ideation.}\\
When asked to describe their current ideation strategies, designers mentioned sketching, whiteboarding, and looking for examples to ideate new layouts. 13 designers mentioned simply placing elements on a design tool's canvas and moving them around to try to generate new ideas. After using Scout, 14 designers said they would use Scout to quickly ideate layouts. When comparing their approaches to ideating alternatives with Scout versus the Baseline, 6 designers mentioned struggling to think of ideas in the Baseline task, like P5:
\vspace{-4pt}
\quotateblock{}{It was definitely more time consuming because I wanted to see a bunch of different things upfront, just to see if different concepts would even work...[P5 describes different ways they moved the elements around the screen.] It would have been nice to quickly see that, like, I didn't want every [element] up there [top of screen], I just wanted profile picture, name and title.}
\vspace{-1pt}

In contrast, P4 pointed out how much easier Scout made it to come up with alternatives:
\vspace{-4pt}
\quotateblock{}{It was way quicker for me to come up with these three [Scout designs]. What I struggled with the most on the first [Baseline task] was really brainstorming and ideating, these sort of different variations. Scout made brainstorming a much easier process.}
\vspace{-1pt}

When asked to describe how Scout might fit into their design process, P6 replied that they might use it to see alternatives that already contained their elements, rather than needing to imagine them based on other examples:
\vspace{-4pt}
\quotateblock{}{Instead of searching on the Internet for alternative layouts or existing things that are out there, [Scout] just makes it easy with what you already have. You already see what [the layout] could look like with the information that you have, and not other information.}
\vspace{-1pt}


\section{Related Work}
\squeeze{Scout is inspired by past systems for interacting with alternatives. DesignScape \cite{ODonovan2015DesignScape} provides alternative suggestions for graphic design layouts using an energy-based model based on design principles.} Sketchsplore \cite{Todi2016Sketchsplore} is an interface sketching tool that provides alternatives generated by human performance models. Scout improves upon these systems by letting designers give direct feedback on attributes of alternatives and by letting designers create high-level constraints on the semantics and emphasis of their interfaces. 

Designers frequently explore alternatives by looking for examples \cite{herring2009examples, lee2010designing}. D.Tour lets designers search for examples by color and style, but not adapt them into their own designs. Other example exploration tools \cite{chang2012webcrystal, lee2010designing} let designers both search for examples and extract styles \cite{chang2012webcrystal, lee2010designing}, copy elements from examples \cite{hartmann2007dmix}, or transform a layout into the content and style of another \cite{kumar2011bricolage}. Rewire~\cite{swearngin2018rewire} lets designers convert examples into editable vectorized mockups. While adapting elements of examples can be useful, the designer cannot easily see the example design with their own elements. Scout lets designers quickly visualize alternatives with their own elements without needing to rearrange or restyle examples. 

Systems for creating and managing alternatives have been created for 2D graphic designs, 2D interfaces, and 3D modeling. For 2D interfaces, Juxtapose \cite{hartmann2008juxtapose} supports simultaneous editing of linked alternatives. Subjunctive interfaces \cite{lunzer2008subjunctive} lets a person simultaneously manage alternatives by editing parametric models. In Parallel Paths \cite{terry2004variation}, people can create alternatives by branching from an initial design. Unlike these systems, Scout requires specifying only \mbox{high-level} semantics (e.g., emphasis), does not require an initial design, and focuses on early ideation support. 
For 2D generative designs, GEM-NI \cite{zaman2015gemni} supports parallel creation and exploration of alternatives. For 3D designs, Dream Sketch \cite{kazi2017dreamsketch} and Dream Lens \cite{matejka2018dream} enable exploration of large-scale generative design alternatives through sketching \cite{kazi2017dreamsketch} or through interface tools for selecting, filtering, and visualizing parameters \cite{matejka2018dream}. Scout supports a similar scope of capabilities (i.e.,~generating, viewing, and comparing alternatives), but focuses on support for exploring 2D interface alternatives.    

To create alternatives, Scout systematically modifies design variables (e.g., arrangement, alignment). This concept is like Parameter Spectrum's approach \cite{terry2002sideviews} that previews alternatives from a range of parameter values. Juxtapose \cite{hartmann2008juxtapose} extends this to interface design by enabling the creation of parallel alternatives through code-based tuning of design parameters. 
\squeeze{Scout does not expose these parameters to designers, however, it would be possible to make these customizable.} 

\squeeze{Model-based user interfaces \cite{zanden1990jade, sukaviriya1993UIDE, szekely1993modelbased, lin2008damask, nichols2004smarttemplates} let designers specify a \mbox{high-level} model and  generate alternatives maintaining the model.} Smart Templates \cite{nichols2004smarttemplates} and Damask \cite{lin2008damask} use models to maintain interface conventions across platforms. Scout similarly maintains high-level constraints across alternatives. Rather than creating templates or patterns, Scout requires defining only high-level constraints between elements, which can enable it to generate many more alternatives. Scout's approach is conceptually closer to \mbox{previously-discussed} generative design approaches \cite{matejka2018dream, kazi2017dreamsketch} or similar approaches in data visualization \cite{moritz2019formalizing}, yet focuses on 2D interface design. 


\squeeze{Scout enables rapid generation of alternatives through constraint solving techniques \cite{zanden1991lapidary,karsenty1993rockit,badros1999constraint, borning2000constraintlayout, Xu2014beautification,Zeidler2013auckland}.} Many past \mbox{constraint-based} layout tools focus primarily on creating a single design. Scout instead can aid ideation by generating many alternatives.  Scout's approach is like that of PBM \cite{hottelier2014pbm}, which exploits constraint ambiguities to explore alternative data visualizations, or Supple's \cite{gajos2007supple} generation of interfaces customized to motor and vision abilities. \squeeze{Zeidler et. al. \cite{zeidler2017Alternates} and Jiang et. al. \cite{jiang2019orc} apply constraints to generate layouts adapted to alternate screen dimensions or orientations.}

\squeeze{Machine learning has also been applied to explore alternatives by transforming the content of an interface into the style and layout of another \cite{kumar2011bricolage}}. LayoutGAN \cite{li2018layoutgan} synthesizes alternative layouts with a generative adversarial network based on modeling geometric relations of 2D elements. Scout's use of constraint solving, rather than machine learning, gives it direct control over the attributes of layouts that the generation algorithms explore. Scout can also generate reasonable alternatives without requiring a design dataset. 
\section{DISCUSSION AND CONCLUSION}
Scout can enhance designer ideation by helping rapidly visualize many layouts through mixed-initiative interaction with high-level constraints and feedback on alternatives. Our evaluation found Scout can aid designers in creating layout ideas they do not believe they would have otherwise thought of, can help designers avoid developing too early of a focus on a single design, and can help designers consider layouts different from established patterns. Scout designs were also more spatially diverse both within and across designers.

\squeeze{Although not statistically significant, our quality analysis found Scout designs were awarded slightly lower overall quality scores by expert designers. This suggests opportunities to improve Scout in terms of balance, emphasis, and alignment. However, participants also had access to the functionality of Adobe XD to refine their designs after Scout ideation (i.e., the same tool used in the baseline). Any difference in quality may therefore be due to a lack of time. Scout required time for specifying elements as well as their grouping, ordering, and emphasis, which may have left less time for refinement of designs at the end of the task. Future work could explore integrating capabilities developed in Scout as a feature in an existing design tool (e.g., Adobe XD), such that elements, grouping, ordering, and emphasis could be inferred from an existing layout to generate new alternatives.}
 
\squeeze{Scout points to a new approach to using constraints to support ideation and presents new techniques for providing feedback to systems applying constraint solving. Future systems can explore: (1) formalizations of interface design principles into tools that help designers apply those principles, especially when supporting novice designers, (2) scaling interactive constraint solving to larger interfaces (e.g., webpages), and (3) defining more layout variables and constraints to enable systems like Scout to explore larger and higher-quality spaces of alternatives.} 

\subsection{Acknowledgements}
This work was supported by the National Science Foundation through awards DGE-1256082, DGE-1762114, CCF-1703304, CCF-1836813, IIS-1702751, and IIS-1735123 and by a Google gift.  

\bibliographystyle{SIGCHI-Reference-Format}
\bibliography{references}

\end{document}